\documentclass[a4paper,11pt]{article}

\usepackage{amsmath,amsthm,amsfonts,amssymb,stmaryrd,fancyhdr,mathtools,latexsym}
\usepackage{float}
\usepackage{fullpage}
\usepackage{authblk}
\usepackage[english]{babel}

\title{Spectral flow and conformal blocks in $AdS_3$}
\author[1]{Yago~Cagnacci\thanks{yago@iafe.uba.ar}}
\author[1,2]{\,Sergio~M.~Iguri\thanks{siguri@iafe.uba.ar}}

\affil[1]{\,Instituto de Astronom\'{\i}a y F\'{\i}sica del Espacio (IAFE, CONICET-UBA)

Casilla de Correo 67, Sucursal 28, 1428 Buenos Aires, Argentina}

\affil[2]{\,Institut de Physique Th\'eorique, CEA/Saclay

91191 Gif-sur-Yvette Cedex, France}

\begin{document}

\maketitle

\begin{abstract}

In this article we investigate the structure of the four-point functions of the $AdS_3$-WZNW model. We consider the integral expression for the unflowed four-point correlator involving at least one state in the discrete part of the spectrum derived by analytic continuation from the $H_3^+$-WZNW model and we show that the conformal blocks can be obtained from those with an extremal-weight state by means of an intertwining operator. We adapt the procedure for dealing with correlators with a single unit of spectral flow charge and we get a factorized integral expression for the corresponding four-point function. We finally transform the formulas back to the space-time picture.

\end{abstract}

\bigskip

\section{Introduction}

The $AdS_3$-WZNW model has proven to be useful in many areas of theoretical physics, ranging from gravity and string theory \cite{Seiberg:1990eb,Witten:1991yr,Banados:1992wn,Mukhi:1993zb,Ghoshal:1995wm,Ooguri:1995wj,Giveon:1998ns,Kutasov:1999xu} to condensed-matter physics \cite{Zirnbauer:1999ua,Bhaseen:1999nm,Kogan:1999hz}. It became particularly popular after the AdS/CFT correspondence \cite{Maldacena:1997re,Witten:1998qj,Aharony:1999ti}, as it describes the worldsheet of a string propagating in an Anti-de Sitter space-time with a background NS-NS $2$-form $B$ field. So far, it is one of the few schemes in which Maldacena's conjecture can be studied beyond the supergravity approximation. The interest in the $AdS_3$-WZNW and other models having a $SL(2,\mathbb R) \times SL(2,\mathbb R)$ global symmetry has been recently renewed within the context of integrability \cite{Beisert:2010jr,OhlssonSax:2011ms,Cagnazzo:2012se,Sfondrini:2014via} and $AdS_3$ gravity \cite{Fjelstad:2011sj,Kirsch:2011na,Polchinski:2012nh,Kim:2015bba,Giveon:2015cma}.

In a WZNW model with a compact underlying symmetry group, the spectrum is built upon representations of the zero-modes algebra \cite{Gepner:1986wi}. Vectors in these modules are assumed to be annihilated by all strictly positive frequency modes and, afterwards, representations of the full current algebra are generated by acting with the modes of negative degree. Unitarity further constrains the space of physical states. For the $AdS_3$-WZNW model, being a non-rational CFT, this procedure does not guarantee the absence of negative-norm states in the spectrum of the string and a bound in the upper value of the spin was suggested to be necessary \cite{Balog:1988jb,Evans:1998qu}, implying a coupling independent restriction on the masses of the physical configurations that originally raised doubts about whether a no-ghost theorem could finally be proven in this case. In addition, the spectrum generated that way gives no account of finite energy states corresponding to strings stretched closed to the boundary of $AdS_3$ \cite{Maldacena:1998uz,Seiberg:1999xz}.

A solution for both issues was presented in \cite{Maldacena:2000hw}. The space of states was enlarged in order to include representations with energy unbounded below by means of the so-called spectral flow automorphisms and a no-ghost theorem based on this spectrum was proven. In \cite{Maldacena:2000kv} the proposal was verified by computing the modular invariant one-loop partition function for Euclidean BTZ black hole backgrounds, more modular properties of the $AdS_3$-WZNW model being further studied in \cite{Israel:2003ry,Baron:2010vf}.

The spectral flow automorphisms are morphisms of the current algebra labelled by an integer number $\omega$ that can be roughly related to the amount of winding of the string in the angular direction of $AdS_3$. For WZNW models based on compact Lie groups, the spectral flow maps primary states of one representation into descendants of another and, therefore, no new configurations are created. However, in a non-rational CFT the spectral flow operation, when applied on an arbitrary module, leads, in general, to a nonequivalent one generated by infinitely many affine quasi-primary fields having energies unbounded from below, thus giving account of long string states.
 
Correlation functions involving only primary fields are typically assumed to follow by analytic continuation of some related functions in the Euclidean counterpart of the theory, {\em i.e.}, the $H_3^+$-WZNW model \cite{Teschner:1997ft,Teschner:1999ug}. These unflowed correlators can be written either in the space-time picture \cite{Maldacena:2001km,Cagnacci:2013ufa}, the vertex fields being parametrized in this representation by the coordinates of the dual CFT target space, or in the so-called $m$-basis \cite{Hosomichi:2001fm,Giveon:2001up,Satoh:2001bi,Minces:2007td}, a basis in which the Cartan generator of $SL(2,\mathbb R)$ is diagonal. Many of them can be computed directly in the Lorentzian model using free fields methods, as in \cite{Becker:1993at,Hosomichi:2000bm,Iguri:2009cf}. It should be mentioned, however, that the contact between the $H_3^+$ and the $AdS_3$-WZNW models is still a matter of discussion \cite{Fjelstad:2011sj,Baron:2008qf,Ribault:2009ui}, the analytic continuation of the corresponding correlators being both a conceptually and a technically difficult task.

When spectral flowed states are also considered, the computation of correlation functions turns out to be even more complicated. The standard strategy for doing so was developed in \cite{Fateev} based on the parafermionic representation. Starting with a correlator with all of their states unflowed, it requires the insertion of an additional vertex, the so-called spectral flow operator, for each unit of spectral flow involved, followed by a transformation to the $m$-basis, an ad-hoc process for removing the dependence of the resulting expression on the unphysical points in which the spectral flow fields were inserted and, if needed, a final transformation back to the space-time basis. Dealing with a large number of insertions makes this method useless when trying to determine correlation functions with more than three physical states. The computation, both in the space-time picture and in the $m$-basis, of two and three-point functions with nontrivial total spectral flow number can be found in \cite{Maldacena:2001km,Cagnacci:2013ufa}. Similar correlators were computed in \cite{Giribet:2001ft,Giribet:2006eb,Iguri:2007af} within the free fields approach. Some results concerning flowed amplitudes with more insertions can be found in \cite{Minces:2005nb,Giribet:2011xf}.

The aim of this paper is to further continue the study of correlation functions in the $AdS_3$-WZNW model. In particular, we analyze the structure of four-point functions with one state lying in the discrete part of the spectrum, both in the unflowed sector and in the case of a single unit of total spectral flow charge. The method we employ is similar to the one used in \cite{Satoh:2001bi,Becker:1993at,Iguri:2007af} when computing three-point functions, as it relies in the knowledge of the correlator when one field is a lowest or a highest-weight state and it relaxes the extremal-weight state condition by exploiting the invariance of a trace of a product under cyclic permutations of the corresponding factors. Let us recall that in \cite{Iguri:2009cf} an integral expression generalizing the classical Selberg integral was found for the one extremal-weight state case using the free fields formalism and this is, so far, the only formula for the $AdS_3$ conformal blocks in the $m$-basis available in the literature.

The paper is organized as follows. In the next section we review the $H_3^+$ and the $AdS_3$-WZNW models. We discuss the similarities and differences between their spectra, the role played by a spectral flow automorphism within this context and the way it enters the computation of correlation functions. In Section \ref{sec3} we study the unflowed four-point function with one state in the discrete spectrum of the theory. We show that the corresponding conformal blocks can be derived from those involving a lowest or a highest-weight state. As we have mentioned before, our method resembles the one already employed in \cite{Satoh:2001bi,Becker:1993at,Iguri:2007af} for determining some of the structure constants of the model. These techniques are afterwards adapted in order to deal with correlators in nontrivial spectral flow sectors, and then they are used in Section \ref{sec4} for obtaining a factorized integral expression for the four-point function with one discrete state and a single unit of spectral flow charge. The conformal blocks in the extremal-weight case are determined using the isomorphism between the unflowed discrete series and the images of certain representations under the action of a spectral flow automorphism. In Section \ref{sec5} we transform these formulas back to the space-time picture and, in a final section, we present our conclusions.

\section{Review of the $H_3^+$ and $AdS_3$-WZNW models}
\label{sec2}

The action of the nonlinear sigma model whose target space is the hyperbolic space $H_3^+$ can be expressed in terms of the Poincar\'e coordinates $(\phi,u,\bar u)$ as
\begin{equation}
\label{action}
\mathcal S = k \int d^2z \left( \partial \phi \bar \partial \phi + e^{2\phi} \partial \bar u \bar \partial u\right),
\end{equation}
where $k$ is the level of the model, $z$ is the complex coordinate of the worldsheet, $\partial$ stands for $\partial/\partial z$ and the bar indicates, as usual, complex conjugation. It has a set of conserved holomorphic and antiholomorphic currents that generate two commuting isomorphic $\mathfrak{sl}_2$-current algebras. Their modes, $J^a_n$ and $\bar{J}^a_n$, with $a=3,\pm$ and $n \in \mathbb Z$ satisfy the following commutation relations,
\begin{equation}
\left[J^3_n,J^3_m\right]=-\frac{1}{2}kn\delta_{n+m,0},
\end{equation}
\begin{equation}
\left[J^3_n,J^{\pm}_m\right]=\pm J^{\pm}_{n+m},
\end{equation}
\begin{equation}
\left[J^-_n,J^+_m\right]=2 J^3_{n+m} + kn\delta_{n+m,0},
\end{equation}
and analogous expressions hold for the antiholomorphic modes. Associated with these algebras there are two commuting Virasoro algebras constructed according to the Sugawara procedure. Their generators are given by
\begin{equation}
\label{Lm}
L_m=\frac{1}{k-2} \sum_{n=1}^{\infty} \left(J^{+}_{m-n}J^{-}_{n}+J^{-}_{m-n}J^{+}_{n} -2 J^3_{m-n}J^3_{n}\right),
\end{equation}
for $m\ne 0$, and
\begin{equation}
\label{L0}
L_0=\frac{1}{k-2} \left[ \frac{1}{2}J^{+}_0J^{-}_0+\frac{1}{2}J^{-}_0J^{+}_0-\left(J^3_0\right)^2 + \sum_{n=1}^{\infty} \left(J^{+}_{-n}J^{-}_{n}+J^{-}_{-n}J^{+}_{n}+ -2 J^3_{-n}J^3_{n}\right)\right],
\end{equation}
and the same for the antiholomorphic generators $\bar L_n$. Both copies of the Virasoro algebra share the same central charge, given in terms of the level by
\begin{equation}
c=\frac{3k}{k-2}.
\end{equation} 

The spectrum of the model is spanned as a direct integral of certain irreducible modules of the current algebra \cite{Teschner:1997ft,Teschner:1999ug}. These modules are freely generated by acting with the negative triangular part of the current algebra, namely, with the generators $J^a_n$ and $\bar{J}^a_n$ with $n<0$, on the representations of the zero-modes corresponding to the principal series of $SL(2,\mathbb C)$. The positive triangular part is assumed to act trivially on these zero-modes representations. Consequently, the space of states is parametrized by a single number, the same that classifies the principal series of $SL(2,\mathbb C)$ inducing their irreducible components. We shall refer to this number as the spin and we will denote it by $j$. Let us recall that $j$ lies in the half-axis $\mathcal C^+ =\{ -1/2 + i\lambda:\lambda \in \mathbb R_{>0}\}$.

A concrete realization of the spectrum can be obtained by means of affine primary operators $\Phi_j(x|z)$ having the following OPE with the currents,
\begin{equation}
\label{ope}
J^a(z)\Phi_j(x|w) = \frac{1}{z-w} D^a_j \Phi_j(x|w),
\end{equation}
for $a=3,\pm$, where
\begin{equation}
D^-_j = -\partial_x,
\end{equation}
\begin{equation}
D^3_j = -x\partial_x+j,
\end{equation}
\begin{equation}
D^+_j = -x^2\partial_x+2jx,
\end{equation}
and  analogously for $\bar{J}^a(z)$ with the complex conjugate differential operators. The label $x$ appears as a complex coordinate of the boundary of $H_3^+$, which becomes the target space of the dual two-dimensional CFT \cite{Giveon:1998ns,Kutasov:1999xu}. We shall therefore refer to this realization as the space-time picture. The operators $\Phi_j(x|z)$ are also primary fields for the Sugawara-Virasoro algebra with conformal dimensions given by
\begin{equation}
\Delta_j=-\frac{j(1+j)}{k-2}.
\end{equation}

Correlation functions on the sphere involving $N$ of these primary operators will be denoted by $\Phi_N(J|X|Z)$, namely,
\begin{equation}
\Phi_N(J|X|Z)=\left\langle \prod_{i=1}^N \Phi_{j_i}(x_i|z_i) \right\rangle,
\end{equation}
where we have introduced $J=(j_1,\dots,j_N)$, $X=(x_1,\dots,x_N)$ and $Z=(z_1,\dots,z_N)$. The conformal weights will also be collectively referred as $\Delta=(\Delta_{j_1},\dots,\Delta_{j_N})$.

The invariance of these correlators under the symmetries generated by $J^a_0$, $\bar{J}^a_0$, $a=3,\pm$, and $L_n$, $\bar{L}_n$, $n = 0,\pm 1$, fully determines the functional dependence of the two and the three-point functions on both the worldsheet and the space-time coordinates, leaving only some constants depending on the kinematical configurations to be further established \cite{Teschner:1997ft,Teschner:1999ug}. When the normalization of the states is fixed as in \cite{Teschner:1999ug}, the propagator turns out to be
\begin{equation}
\label{Euclidean2pt}
\Phi_2(J|X|Z) =  |z_{12}|^{-4\Delta_{j_1}}\left[\delta^2(x_{12})\delta(1+j_1+j_2) +  |x_{12}|^{4j_2} B(j_1)\delta(j_2-j_1) \right],
\end{equation}
where $x_{12}=x_2-x_1$, $z_{12}=z_2-z_1$,
\begin{equation}
B(j) = \frac{\nu^{1+2j}}{\pi b^2} \gamma\left(1+b^2(1+2j)\right),
\end{equation}
\begin{equation}
\nu = \frac{\pi}{b^2} \gamma\left(1-b^2\right),
\end{equation}
with $b^2=(k-2)^{-1}$ and $\gamma(x)=\Gamma(x)/\Gamma(1- \bar x)$. The three-point function takes the form
\begin{equation}
\label{3ptbasex}
\Phi_3(J|X|Z)= |C(\Delta|Z)|^{-2}|C(J|X)|^2 D(J) ,
\end{equation}
where the structure constants $D(J)$ and the Clebsch-Gordan-like coefficients $C(J|X)$ are explicitly given by
\begin{equation}
\label{structureconst}
D(J) = \frac{G(1+j_1+j_2+j_3)}{\nu^{-1-j_1-j_2-j_3} G_0}\prod_{\sigma}\frac{G(j_{\sigma})}{G(1+2j_{\sigma_1})},
\end{equation}
and
\begin{equation}
\label{clebsch}
C(J|X)=\prod_{\sigma} x_{\sigma_1\sigma_2}^{j_{\sigma}}.
\end{equation}
In (\ref{structureconst}) and (\ref{clebsch}) the products run over all cyclic permutations of the labels and $j_{\sigma}= j_{\sigma_1}+j_{\sigma_2}-j_{\sigma_3}$. The special function $G(j)$ is defined \cite{Zamolodchikov:1995aa} in terms of the Barnes double gamma function $\Gamma_2$ as follows,
\begin{equation}
\label{G}
G(j) = b^{-bj\left(b+b^{-1}+bj\right)}\Gamma_2(-bj|b,b^{-1})\Gamma_2(b+b^{-1}+bj|b,b^{-1}),
\end{equation}
and
\begin{equation}
G_0=-2\pi^2\gamma(1+b^2)G(-1).
\end{equation}
The function $C(\Delta|Z)$ is
\begin{equation}
C(\Delta|Z)=\prod_{\sigma} z_{\sigma_1\sigma_2}^{\Delta_{\sigma}},
\end{equation}
where $\Delta_{\sigma}= \Delta_{j_{\sigma_1}}+\Delta_{j_{\sigma_2}}-\Delta_{j_{\sigma_3}}$. The dependence of the three-point function on the boundary coordinates is fixed by $SL(2,\mathbb C)$ invariance as long as no $j_{\sigma}$ equals a negative integer for any cyclic permutation $\sigma$. 

The action of the $AdS_3$-WZNW model is formally related to the action (\ref{action}) of the $H_3^+$-WZNW model by a Wick rotation of its field variables and, therefore, a close connection between both theories is usually assumed. This contact, however, turns out to be a rather nontrivial one since it is expected to give account of the very distinctive properties these models show. One example is given by their spectra; while the space of states of the $H_3^+$-WZNW model only involves a continuous series of representations of the symmetry algebra, its Lorentzian counterpart is fairly more complicated as it incorporates also a discrete series and their images under the so-called spectral flow automorphisms. These morphisms constitute a key ingredient for generating a consistent spectrum for the $AdS_3$-WZNW model but, in principle, they play no role in the Euclidean theory. Given $\omega \in \mathbb Z$, the corresponding spectral flow automorphism is defined by
\begin{equation}
\label{spectral1}
J^3_n \rightarrow J^3_n - \frac{k}{2} \omega \delta_{n,0},
\end{equation}
\begin{equation}
\label{spectral2}
J^{\pm}_n \rightarrow  J^{\pm}_{n \pm \omega},
\end{equation}
and similarly for the antiholomorphic modes and, as it can be straightforwardly deduced from (\ref{Lm})-(\ref{L0}), it maps the Sugawara-Virasoro algebra into another conformal realization generated, this time, by
\begin{equation}
\label{flowVir}
L_n \rightarrow  L_n + \omega J^3_n - \frac{k}{4}\omega^2\delta_{n,0},
\end{equation}
for $n\in\mathbb Z$, and correspondingly for the $\bar L_n$. Unlike in rational models, which have underlying compact group symmetries, the spectral flow automorphisms generally give rise to nonequivalent representations when acting on a current module.

Let us be more specific concerning the space of physical states of the $AdS_3$-WZNW model. As it was conjectured in \cite{Maldacena:2000hw}, it consists of two families of representations of the current algebra, the first one composed by the modules induced by the principal continuous representations of the universal cover of $SL(2,\mathbb R)$ and their spectral flow images, and the other composed by those induced by the principal discrete series and their spectral flow images, all of them with the same quantum numbers for the left and the right sectors. For the continuous series these numbers are the spin $j$, lying in $\mathcal C^+$ as for the $H_3^+$-WZNW model, and a real parameter $\alpha\in [0,1)$. The discrete series are labelled by a single number, also denoted by $j$, that must be in the real half-axis $(-\infty,-1/2)$ once unitarity is required. Every spectral flow image of a representation induced by a lowest-weight discrete series with spin $j$ is isomorphic to one built upon a highest-weight discrete series with a spin $j'=-k/2-j$ and an additional unit of flow. This module isomorphism, that will be referred in the sequel as the series identification, restricts the discrete representations allowed in the spectrum to be either those induced by the lowest or by the highest-weight series, while constraining the range of values of the spin to the real interval
\begin{equation}
-\frac{k-1}{2}<j<-\frac{1}{2}.
\end{equation}

A suitable realization of the spectrum can be obtained by means of vertex operators in the so-called $m$-basis, where the label $m$ is introduced in order to keep track of the quantum number associated with the eigenvalue of $J_0^3$ in the unflowed frame. This basis is the best suited to a Wick rotation from the $H_3^+$-WZNW model as well as to further include a spectral flow charge when computing correlation functions. We shall denote the affine primary fields realizing the unflowed spectrum by $\Phi_j(m|z)$ and their images under a spectral flow automorphism or, for short, the flowed vertex operators, by $\Phi^{\omega}_j(m|z)$. When dealing with states in the $\omega=\mp 1$ sector we shall adopt another notation and we will write $\widehat{\Phi}_j(m|z)$ instead. A dependence on an antiholomorphic label $\bar m$ is implicitly assumed, with $m-\bar m \in \mathbb Z$ and $m+\bar m \in \mathbb R$.  
The OPE of these fields with the currents and their conformal weights are given by
\begin{equation}
\label{OPE1}
J^{3}(z)\Phi^{\omega}_j(m|w) = \frac{m+k\omega/2}{z-w}\Phi^{\omega}_j(m|w),
\end{equation}
\begin{equation}
\label{OPE2}
 J^{\pm}(z)\Phi^{\omega}_j(m|w) = \frac{\mp j + m}{(z-w)^{1 \pm \omega}}\Phi^{\omega}_j(m\pm 1|w),
\end{equation}
\begin{equation}
\label{confweiw}
\Delta_{jm}^{\omega}=\Delta_j -\omega m - \frac{k}{4} \omega^2,
\end{equation}
and analogously for the antiholomorphic counterparts. When $\omega=\mp 1$ we shall write $\widehat \Delta_{jm}$ instead of $\Delta_{jm}^{\omega=\mp 1}$.

The $N$-point correlation functions in the $AdS_3$-WZNW model involving these spectral flowed fields will generally be denoted by
\begin{equation}
\Phi_N(J|\Omega|M|Z)=\left\langle \prod_{i=1}^N \Phi^{\omega_i}_{j_i}(m_i|z_i) \right\rangle,
\end{equation}
with $J$ and $Z$ as before, $M=(m_1,\dots,m_N)$ and $\Omega=(\omega_1,\dots,\omega_N)$. A salient feature of these correlators, with the only exception of the two-point function, is that they can exhibit a violation of the spectral flow number conservation, namely, it could happen for $N \ge 3$ that the total amount of flow, $\omega=\sum_{i=1}^{N}\omega_i$, does not vanish. This non-conservation of the total spectral flow number is regulated by a selection rule that can be stated as follows,
\begin{equation}
\label{violation1}
-N+\delta \le \omega \le N_c-\delta,
\end{equation}
where $N_c$ is the number of vertex operators associated with states lying in the continuous spectrum, $\delta=1$ when $N_c=0$ and $\delta=2$ otherwise. It can be easily seen, by virtue of the series identification, that (\ref{violation1}) states that 
an $N$-point amplitude can reach a maximal violation of $N-2$ units of spectral flow. Another relevant aspect of the $AdS_3$-WZNW model correlators is that two amplitudes with the same value of $\omega$  only differ in a factor that adjusts the overall worldsheet dependence. Indeed, it can be deduced from \cite{Fateev} that $F_N(\Omega|M|Z) \Phi_N(J|\Omega|M|Z)$, with
\begin{equation}
F_N(\Omega|M|Z)=\left|\prod_{i<j}^N (z_i-z_j)^{m_i\omega_j+m_j\omega_i+\omega_i\omega_jk/2}\right|^2,
\end{equation}
remains the same for every spectral flow assignment adding up to $\omega$. Consequently, for studying an $N$-point correlation function there is no loss of generality if we assume that only the vertex in the first insertion can be flowed, with an amount of spectral flow running from $0$ to a limiting absolute value of $N-2$. We shall do so in the rest of the paper. Unflowed $N$-point amplitudes will accordingly be assumed to have $\omega_i=0$ for $i=1,\dots,N$, and they will be denoted by $\Phi_N(J|M|Z)$. Correlators with one unit of spectral flow, either negative or positive, {\em i.e.}, with $\Omega=(\mp 1,0,\dots,0)$, will be denoted by $\widehat{\Phi}_N(J|M|Z)$ and the collective variable for the conformal weights will be $\widehat{\Delta}=(\widehat{\Delta}_{j_1m_1},\Delta_{j_2},\dots,\Delta_{j_N})$. We shall not consider correlation functions violating the spectral flow number conservation in more than a single unit.

Vertex operators satisfying (\ref{OPE1})-(\ref{confweiw}) in the unflowed sector can be obtained from those of the $H_3^+$-WZNW model by means of the following Mellin-like transform
\begin{equation}
\label{fourier}
 \Phi_{j}(m|z) = \int_{\mathbb C} d^2x x^{j+m} \bar x^{j+ \bar m} \Phi_{-1-j}(x|z),
\end{equation}
and, consequently, a correlation function $\Phi_N(J|M|Z)$ involving these fields is expected to be obtained after using the transformation (\ref{fourier}) on each insertion in the corresponding Euclidean correlator $\Phi_N(J|X|Z)$. Note, however, that in the $H_3^+$-WZNW model, $m+\bar m$ is set as a purely imaginary number, unlike the $AdS_3$-WZNW model, in which $m+\bar m$ is real. It follows that a Wick rotation of $m_i+\bar m_i$, $i=1,\dots,N$, should be performed in order to get the right expression for the Lorentzian $N$-point correlation function. Moreover, the spin, which in the $H_3^+$-WZNW model is restricted to lie in $\mathcal C^+$, must be promoted to a real parameter if (\ref{fourier}) is intended to describe states in the discrete spectrum of the $AdS_3$-WZNW model. Therefore, a well defined analytical continuation on $j_i$, $i=1,\dots,N$, beyond $\mathcal C^+$ should also be argued for obtaining $\Phi_N(J|M|Z)$. Although a rigorous justification of all these assumptions is still lacking, it was under these hypotheses that most of the results concerning the $AdS_3$-WZNW model were hitherto reached. We shall therefore adopt this approach throughout the rest of the paper.

The two-point function, when computed from (\ref{Euclidean2pt}) using (\ref{fourier}), leads to
\begin{equation}
\label{2ptm}
\Phi_2(J|M|Z) = \left|z_{12}\right|^{-4\Delta_{j_1}}\delta[M] \left[\delta(1+j_1+j_2) + c^{}_{-1-j_1,m_1} B(-1-j_1) \delta(j_2-j_1) \right] ,
\end{equation}
where
\begin{equation}
 c_{j m}=\frac{\pi}{\gamma(-2j)}\frac{\gamma(-j+m)}{\gamma(1+j+m)},
\end{equation}
and $\delta[M]$ is an abbreviation for $\delta^{2}(m_1+m_2)$, with
\begin{equation}
\label{deltacomp}
 \delta^2(m)=\int_{\mathbb C} d^2x x^{m-1} \bar x^{\bar m-1} = 4\pi^2 \delta(m+\bar m)\delta_{m-\bar m,0}.
\end{equation}
The unflowed three-point function is 
\begin{equation}
\label{3ptm}
\Phi_3(J|M|Z)= |C(\Delta|Z)|^{-2}\delta[M]W(J|M)  D(-1-J) ,
\end{equation}
with $W(J|M)$ defined as
\begin{equation}
\label{satoh}
W(J|M) = \int_{\mathbb C} d^2x_1 d^2x_2 \prod_{\sigma} x_{\sigma_1}^{j_{\sigma_1}+m_{\sigma_1}} \bar x_{\sigma_1}^{j_{\sigma_1}+ \bar m_{\sigma_1}} \left|x_{\sigma_1\sigma_2}\right|^{-2-2j_{\sigma}},
\end{equation}
where $x_3,\bar x_3=1$. This integral was explicitly computed in terms of certain generalized hypergeometric functions in \cite{Satoh:2001bi,Hosomichi:2000bm}. All these correlators were independently determined within the free fields formalism in \cite{Becker:1993at,Hosomichi:2000bm,Giribet:2001ft}.

When the spectral flow enters the play, the computation of amplitudes is fairly more involved. Indeed, the standard procedure for introducing the violation of the spectral flow number conservation demands the insertion of, at least, one additional vertex, the so-called spectral flow operator, for each unit of flow violation. This method, developed in \cite{Fateev}, was exploited in \cite{Maldacena:2001km} for obtaining the following expression for the spectral flowed three-point function,
\begin{eqnarray}
\label{3ptmw}
&& \widehat{\Phi}_3(J|M|Z) = |C(\widehat{\Delta}|Z)|^{-2}\delta[M\pm k/2]  \widehat{W}(J|M)\widehat{D}(-1-J) ,
\end{eqnarray}
with
\begin{equation}
\widehat{D}(J) = B(j_1) \gamma(1-j'_1+j_2+j_3) D(J'),
\end{equation}
where, as before, $j'_1=-k/2-j_1$ and $J'=(j'_1,j_2,j_3)$, and
\begin{equation}
\label{Wtilde}
\widehat{W}(J|M) = 
(-1)^{m_3-\bar m_3} \prod_{i=1}^3 \gamma(1+j_i \pm m_i).
\end{equation}
An independent computation of this correlation function was performed in \cite{Iguri:2007af} using free fields methods.

\section{Unflowed conformal blocks in $AdS_3$}
\label{sec3}

An integral factorized expression for the four-point function of the $H_3^+$-WZNW model was introduced and intensively studied in \cite{Teschner:1999ug}. It takes the form
\begin{equation}
\label{eucl4pt}
\Phi_4(J|X|Z) = \left| \Xi(\Delta|Z)\right|^{-2} \int_{\mathcal C^+} dj \, \mathcal D_j(J) \mathcal G_j (J|X|z).
\end{equation}
The function $\Xi(\Delta|Z)$ is given by
\begin{equation}
\Xi(\Delta|Z) = z_{43}^{-\Delta_{j_1}-\Delta_{j_2}+\Delta_{j_3}+\Delta_{j_4}} z_{42}^{2\Delta_{j_2}} z_{41}^{\Delta_{j_1}-\Delta_{j_2}-\Delta_{j_3}+\Delta_{j_4}} z_{31}^{\Delta_{j_1}+\Delta_{j_2}+\Delta_{j_3}-\Delta_{j_4}}.
\end{equation}
The factor $\mathcal D_j(J)$ is, as expected, the properly normalized product of two three-point functions involving the intermediate state, namely,
\begin{equation}
\label{product3pt}
\mathcal D_j(J) = D_{12}(j) B(j)^{-1} D_{34}(j),
\end{equation}
where $D_{12}(j)$ is an abbreviation for $D(j_1,j_2,j)$. The so-called non-chiral conformal block $\mathcal G_j (J|X|z)$ is decomposed as
\begin{equation}
\mathcal G_j (J|X|z) = |z|^{2\left(\Delta_j-\Delta_{j_1}-\Delta_{j_2}\right)} \mathcal K_j(J|X|z) G_j (J|X),
\end{equation}
where $z$ is the worldsheet cross-ratio, {\em i.e.}, $z=z_{12}z_{34}/z_{13}z_{24}$, $G_j (J|X)$ is given by
\begin{equation}
\label{intint2}
G_j (J|X) = \frac{(1+2j)^2}{\pi^2} \int_{\mathbb C} d^2x d^2x' C_{12}(j|x) |x-x'|^{-4-4j} C_{34}(j|x'),
\end{equation}
with $C_{12}(j|x)=C(j_1,j_2,j|x_1,x_2,x)$, and the operator $\mathcal K_j(J|X|z)$ has the following factorized form
\begin{equation}
\mathcal K_j(J|X|z) = \mathcal O_j(J|X|z) \overline{\mathcal O_j(J|X|z)},
\end{equation}
where $\mathcal O_j(J|X|z)$ is formally given by the following power series in $z$,
\begin{equation}
\label{expansion}
\mathcal O_j(J|X|z) = \sum_{n=0}^{\infty} z^n \delta_j^{(n)}(J|X).
\end{equation}
In this last expression, $\delta^{(n)}_j(J|X)$ are differential operators containing derivatives of finite order in $x_1, \dots, x_4$. Albeit the explicit formula of theses operators is still unknown, they can be recursively determined by using the Knizhnik-Zamolodchikov equation once $\delta_j^{(0)}(J|X)$ is fixed. In our case we have $\delta_j^{(0)}(J|X)=1$.

By exploiting $SL(2,\mathbb C)$ invariance, the integral (\ref{intint2}) can be simplified. Indeed, one has
\begin{equation}
\label{futGK}
G_j (J|X) = \left| \Xi(J|X) \right|^{2} G_j (J|x),
\end{equation}
where
\begin{equation}
\Xi(J|X) = x_{43}^{-j_{1}-j_{2}+j_{3}+j_{4}} x_{42}^{2j_{2}} x_{41}^{j_{1}-j_{2}-j_{3}+j_{4}} x_{31}^{j_{1}+j_{2}+j_{3}-j_{4}},
\end{equation}
and $G_j (J|x)$ equals (\ref{intint2}) with $x_1=0$, $x_2=x=x_{12}x_{34}/x_{13}x_{24}$, $x_3=1$ and $x_4=\infty$. This integral can be easily carried out, giving
\begin{equation}
\label{Gj}
G_j (J|x)= \left| F_j(J|x) \right|^2 + \lambda_j(J) \left| F_{-1-j}(J|x) \right|^2,
\end{equation}
where
\begin{equation}
\lambda_j(J) = \frac{\gamma(1+j+j_3-j_4)\gamma(1+j-j_3+j_4)}{\gamma(1+2j)\gamma(-j+j_1-j_2)\gamma(-j-j_1+j_2)},
\end{equation}
and $F_j(J|x)$ is
\begin{equation}
\label{hyper}
F_j(J|x) = x^{-j+j_1+j_2} F(-j+j_1-j_2, -j-j_3+j_4;-2j|x).
\end{equation}

Both terms in the right hand side of (\ref{Gj}) are related by a reflection in $j$, {\em i.e.}, $j \leftrightarrow -1-j$. This fact allows to extend the integration in (\ref{eucl4pt}) over the full axis $\mathcal C=\{-1/2+i\lambda:\lambda\in\mathbb R\}$, the resulting expression showing a holomorphically factorized form,
\begin{equation}
\label{eucl4ptfact}
\Phi_4(J|X|Z) = \left| \Xi(\Delta|Z)\right|^{-2} \int_{\mathcal C} dj \, \mathcal D_j(J) \left| \mathcal F_j (J|X|z)\right|^2,
\end{equation}
where the chiral blocks $\mathcal F_j (J|X|z)$ are given by
\begin{equation}
\mathcal F_j (J|X|z) = z^{\Delta_j-\Delta_{j_1}-\Delta_{j_2}} \mathcal O_j(J|X|z) \Xi(J|X) F_j (J|x).
\end{equation}
The $SL(2,\mathbb C)$ invariance also implies that the operator defined as
\begin{equation}
\mathcal O_j (J|x|z) = \Xi(J|X)^{-1} \mathcal O_j(J|X|z) \Xi(J|X),
\end{equation}
actually depends on the cross-ratio $x$, and, therefore, after introducing
\begin{equation}
\mathcal F_j (J|x|z) = z^{\Delta_j-\Delta_{j_1}-\Delta_{j_2}} \mathcal O_j(J|x|z) F_j (J|x),
\end{equation}
we can write
\begin{equation}
\label{eucl4ptfactb}
\Phi_4(J|X|Z)  = \left| \Xi(\Delta|Z)\right|^{-2} \left|\Xi(J|X) \right|^2 \int_{\mathcal C} dj \, \mathcal D_j(J) \left| \mathcal F_j (J|x|z)\right|^2.
\end{equation}

Using expansion (\ref{expansion}), we find
\begin{equation}
\label{expansion2}
\mathcal F_j (J|x|z) = z^{\Delta_j-\Delta_{j_1}-\Delta_{j_2}} x^{-j+j_1+j_2} \sum_{n=0}^{\infty} z^n F^{(n)}_j (J|x).
\end{equation}
where we have introduced
\begin{equation}
F^{(n)}_j (J|x) = \delta^{(n)}_j (J|x) F_j (J|x),
\end{equation}
with
\begin{equation}
\delta^{(n)}_j (J|x) = x^{j-j_1-j_2} \Xi(J|X)^{-1} \delta_j^{(n)}(J|X) \Xi(J|X).
\end{equation}

Both the equation (\ref{eucl4ptfactb}) and the expansion (\ref{expansion2}) were used in \cite{Maldacena:2001km} for studying the four-point function within the context of the AdS/CFT correspondence. Indeed, since $\delta_j^{(0)}(J|X)=1$, it follows that $\delta_j^{(0)}(J|x)=x^{j-j_1-j_2}$ and, therefore, $F^{(0)}_j (J|x)$ equals the hypergeometric function in (\ref{hyper}), in agreement with \cite{Maldacena:2001km}.

A salient feature of having closed the integration contour in (\ref{eucl4ptfact}) is that this formula allows to define a proper meromorphic continuation of the four-point function for generic complex values of $j_1,\dots,j_4$ by exploiting the pole structure of the integrand. For those values of $j_1,\dots,j_4$ lying in the maximal region in which the external spins can be continuously varied in such a way that none of the poles from neither the integrand in the OPE of the first two fields nor the integrand in the OPE of the last two operators giving (\ref{eucl4ptfact}) crosses the integration contour $\mathcal C$, the integral expression for the four-point correlator is retained. This domain turns out to be
\begin{equation}
\label{teschnerdomain1}
\left|\mbox{Re}\left(j_{21}^{\pm}\right)\right|<1/2, \quad \quad j_{21}^+=1+j_1+j_2, \quad j_{21}^-=j_1-j_2,
\end{equation}
\begin{equation}
\label{teschnerdomain2}
\left|\mbox{Re}\left(j_{43}^{\pm}\right)\right|<1/2, \quad \quad j_{43}^+=1+j_3+j_4, \quad j_{43}^-=j_3-j_4.
\end{equation}
Beyond this region, some additional terms coming from the residues of such poles must be taken into account. These terms are associated to contributions of intermediate discrete states. See \cite{Teschner:1999ug,Maldacena:2001km} for more details. For simplicity, in this paper we shall restrict the kinematical parameters to lie in the integral domain (\ref{teschnerdomain1})-(\ref{teschnerdomain2}), so that formulas (\ref{eucl4pt}) and (\ref{eucl4ptfact}) can be interpreted as the unflowed space-time picture four-point function in the $AdS_3$-WZNW model.

As we have already pointed out, in order to further incorporate the spectral flow into the computation we need to transform the expression (\ref{eucl4pt}) to the $m$-basis using (\ref{fourier}). We get
\begin{equation}
\label{integral}
\Phi_4(J|M|Z) =\left| \Xi(\Delta|Z) \right|^{-2} \int_{\mathcal C^+} dj \, |z|^{2\left(\Delta_j-\Delta_{j_1}-\Delta_{j_2}\right)} \mathcal D_j(-1-J) {\mathcal K}_j(J|M|z) G_j (J|M),
\end{equation}
where $\mathcal D_j(-1-J)$ is meant to indicate that the formula (\ref{product3pt}) is evaluated with all the spins reflected and $M=(m_1,m_2,m_3,m_4)$. In this last equation, ${\mathcal K}_j(J|M|z)$ is the transform of the operator $\mathcal K_j(J|X|z)$, while $G_j (J|M)$ is the transform of the function $G_j (J|X)$, namely
\begin{equation}
G_j (J|M) = \int_{\mathbb C^4} \prod_{i=1}^4 \left[d^2x_i \, x_i^{j_i+m_i} \bar x_i^{j_i+\bar m_i}\right] G_j (-1-J|X).
\end{equation}

Eq.~(\ref{integral}) is valid as long as the integration over $\mathcal C$ in (\ref{eucl4pt}) and those over $\mathbb C^4$ coming from (\ref{fourier}) can be interchanged. This fact, far from being a subtlety, concerns the Wick rotation of each $m_i+\bar m_i$, $i=1,\dots,4$, needed for obtaining the Lorentzian correlators from the corresponding correlation functions in the $H_3^+$-WZNW model. Indeed, in order to freely interchange the integration of the spin and the space-time integrations when computing the OPE between two generic primary fields, some additional constraints are necessary to be imposed on $m=m_1+m_2$ and $\bar m=\bar m_1+\bar m_2$, namely, $\max\{\mbox{Re}(m),\mbox{Re}(\bar m)\}>1/2$ and/or $\min\{\mbox{Re}(m),\mbox{Re}(\bar m)\}<-1/2$, depending on which series the states belong to. These conditions guarantee that the singularities of the integrand of the OPE that depend on $m$ and $\bar m$ are well located in the complex $j$-plane. Relaxing any of them would require a proper analytic continuation to give account of the residues of the poles crossing the integration contour, exactly as in the space-time picture. An extensive analysis of these discrete contributions to the $AdS_3$-WZNW model OPE can be found in \cite{Baron:2008qf} and also in \cite{Ribault:2009ui} in the minisuperspace limit. There are, however, certain configurations allowing to neglect any term of the Lorentzian four-point function except the integral (\ref{integral}) without the need of additional constraints on $m$ and $\bar m$.  In this paper we prefer to retain the expression of the four-point correlator the simplest as possible and, therefore, we will consider a concrete case in which this happens, namely,  when at least two external fields, say those at the first and the last insertion points, are assumed to belong to the discrete part of the spectrum. If this condition is relaxed by assuming that the correlator has at least one discrete state, the integral (\ref{integral}) should be redefined to give account of some additional discrete contributions, but aside of this, none of the results in what follows would actually be compromised.

The computation of $G_j (J|M)$ was explicitly performed in \cite{Iguri:2009cf} using free fields methods. An entirely independent check of the result was performed in \cite{Baron:2008qf} by means of the OPE of two primary fields. It is given by
\begin{equation}
\label{minces}
G_j (J|M) = \delta\left[M\right]W_{12}(-1-j) c_{jm}^{-1} W_{34}(-1-j),
\end{equation}
where $W_{12}(j)=W(j_1,j_2,j|m_1,m_2,m)$.

This equation allows to write the following expression for the four-point correlator,
\begin{equation}
\label{eqfinal}
\Phi_4(J|M|Z) =  \left| \Xi(\Delta|Z) \right|^{-2} \int_{\mathcal C^+} dj \, \mathcal D_j(J|M) {\mathcal G}_j (J|M|z),
\end{equation}
with
\begin{equation}
\label{cjm}
\mathcal D_j(J|M) = \mathcal D_j(-1-J) G_j (J|M) = \Phi_{12}(j|-m) \Phi_2(j|m)^{-1} \Phi_{34}(j|m),
\end{equation}
where $\Phi_{12}(j|m) = \Phi_3(j_1,j_2,j|m_1,m_2,m|0,1,\infty)$, $\Phi_2(j|m)=\Phi_2(j,j|m,-m|0,\infty)$ and the conformal blocks are given by
\begin{equation}
\label{confblock}
{\mathcal G}_j (J|M|z) = |z|^{2\left(\Delta_j-\Delta_{j_1}-\Delta_{j_2}\right)} G_j (J|M)^{-1}{\mathcal K}_j(J|M|z)G_j (J|M).
\end{equation}
Notice that ${\mathcal K}_j(J|M|z)\rightarrow 1$ and, therefore, ${\mathcal G}_j (J|M|z) \sim |z|^{2\left(\Delta_j-\Delta_{j_1}-\Delta_{j_2}\right)}$, when $z\rightarrow 0$, as expected.

As for the $H_3^+$-WZNW model, the explicit form of the conformal blocks is not known. However, although the existence of a closed formula for ${\mathcal G}_j (J|M|z)$ is unlikely for the general case, in \cite{Iguri:2009cf} an integral expression generalizing the classical Selberg integral was found when the spins satisfy $j_1+j_2+j_3+j_4+1 \in \mathbb N_0$, as it is usual in the free fields formalism, and the first insertion corresponds to a lowest or a highest-weight state (see formulas (2.23)-(2.31) in \cite{Iguri:2009cf}). The first constraint can eventually be bypassed by a proper, and highly nontrivial, analytic continuation, that is beyond the scope of this paper. The extremal-weight condition, on the other hand, can be relaxed by following a procedure similar to the one used in \cite{Satoh:2001bi,Becker:1993at,Iguri:2007af} for the three-point functions. We shall do so in the rest of this section.

From the Baker-Campbell-Hausdorff formula
\begin{equation}
e^{\alpha J_0^{\pm}} \Phi_{j}(m|z) e^{-\alpha J_0^{\pm}} = e^{\alpha \left[ J_0^{\pm} , ~ \right]} \Phi_{j}(m|z) = \sum_{\lambda=0}^{\infty} \frac{\alpha^{\lambda}}{\lambda!} \left[ J_0^{\pm},\Phi_{j}(m|z)\right]_{\lambda},
\end{equation}
where we have defined, inductively,
\begin{equation}
\left[ J_0^{\pm},\Phi_{j}(m|z)\right]_{0} = \Phi_{j}(m|z),
\end{equation}
\begin{equation}
\left[ J_0^{\pm},\Phi_{j}(m|z)\right]_{\lambda} = \left[ J_0^{\pm},\left[ J_0^{\pm},\Phi_{j}(m|z)\right]_{\lambda-1}\right],
\end{equation}
and Eqs.~(\ref{OPE1})-(\ref{OPE2}), it is straightforward to prove that
\begin{equation}
\label{bb}
e^{\pm\alpha J_0^{\pm}} \Phi_{j}(m|z) e^{\mp\alpha J_0^{\pm}} = \sum_{\lambda=0}^{\infty} \frac{\alpha^{\lambda}}{\lambda!} \frac{\Gamma(-j\pm m+\lambda)}{\Gamma(-j\pm m)} \Phi_{j}(m\pm\lambda|z),
\end{equation}
for any complex value of $\alpha$. This formula can be used in order to relate different correlators by exploiting the invariance of the trace under any cyclic permutation of their arguments. Indeed, we have
\begin{equation}
\label{campbell}
\left\langle e^{\pm\alpha J_0^{\pm}} \Phi_{j_1}(m_1|z_1) e^{\mp\alpha J_0^{\pm}} \prod_{i=2}^N \Phi_{j_i}(m_i|z_i) \right\rangle =  \left\langle \Phi_{j_1}(m_1|z_1) \prod_{i=2}^N e^{\mp\alpha J_0^{\pm}} \Phi_{j_i}(m_i|z_i) e^{\pm\alpha J_0^{\pm}} \right\rangle,
\end{equation}
which after replacing (\ref{bb}) implies
\begin{eqnarray}
\label{coorr}
&& \sum_{\lambda_1=0}^{\infty} \frac{\alpha^{\lambda_1}}{\lambda_1!} \frac{\Gamma(-j_1\pm m_1+\lambda_1)}{\Gamma(-j_1\pm m_1)} \left\langle \Phi_{j_1}(m_1\pm\lambda_1|z_1) \prod_{i=2}^N \Phi_{j_i}(m_i|z_i) \right\rangle \nonumber \\
&& = \sum_{\lambda_1=0}^{\infty} (-\alpha)^{\lambda_1} \sum_{\lambda} \left( \prod_{i=2}^N \frac{\Gamma(-j_i\pm m_i+\lambda_i)}{\lambda_i!\Gamma(-j_i\pm m_i)} \right) \left\langle \Phi_{j_1}(m_1|z_1) \prod_{i=2}^N  \Phi_{j_i}(m_i\pm\lambda_i|z_i) \right\rangle.
\end{eqnarray}
The second sum in the right hand side of (\ref{coorr}) runs over all $(N-1)$-tuples with non-negative integer entries $\lambda=(\lambda_2,\dots,\lambda_N)$ adding up to $\lambda_1$
. Identifying each power of $\alpha$, renaming $m_1 \rightarrow m_1\mp\lambda_1$ and making explicit the antiholomorphic dependence, we get
\begin{equation}
\label{bbfinal2b}
\Phi_N(J|M|Z) =   \mathcal Q^{\lambda_1}_N(J|M|\lambda) \Phi_N(J|M\pm \Lambda|Z),
\end{equation}
where $\Lambda=(-\lambda_1,\lambda)$ and we have introduced
\begin{equation}
\label{bbfinal2bbis}
\mathcal Q^{\lambda_1}_N(J|M|\lambda) =   \left| \frac{(-1)^{\lambda_1}\lambda_1!\Gamma(-j_1\pm m_1-\lambda_1)}{\prod_{i=1}^N \Gamma(-j_i\pm m_i)} \prod_{i=2}^N \frac{1}{\lambda_i!}\Gamma(-j_i\pm m_i+\lambda_i) \right|^2.
\end{equation}
In (\ref{bbfinal2b}), $\lambda$ appears as a multi-index so that Einstein summation convention holds; $\mathcal Q^{\lambda_1}_N(J|M|\lambda)$ is thus realized as an operator acting by contraction. Notice that $\lambda_1$ is a parameter assumed to be fixed while defining (\ref{bbfinal2bbis}).

A useful identity follows from (\ref{bbfinal2b}) and (\ref{bbfinal2bbis}). In (\ref{bbfinal2b}), the contraction of $\lambda$ can be performed as a sum over $\lambda_2$ running from $0$ to $\lambda_1$ followed by a sum over all $(N-2)$-tuples with non-negative integer entries $\widetilde \lambda=(\lambda_3,\dots,\lambda_N)$ adding up to $\lambda_1-\lambda_2$. Using (\ref{bbfinal2bbis}) we obtain
\begin{eqnarray}
\label{identity}
&& \mathcal Q^{\lambda_1}_N(J|M|\lambda) = \frac{(-1)^{\lambda_2}\lambda_1!\Gamma(-j_1\pm m_1-\lambda_1)\Gamma(-j\pm m)}{\Gamma(j_1\pm m_1)\Gamma(j_2\pm m_2)} \times \nonumber \\
&& \sum_{\lambda_2=0}^{\lambda_1} \frac{\Gamma(-j_2\pm m_2+\lambda_2) }{\lambda_2!(\lambda_1-\lambda_2)!\Gamma(-j\pm m-\lambda_1+\lambda_2)} \mathcal Q_{N-1}^{\lambda_1-\lambda_2}(j,j_3,\dots,j_N|m,m_3,\dots,m_N|\widetilde \lambda),
\end{eqnarray}
where we have introduced two new auxiliary variables $j$ and $m$, and there is an implicit sum after the right hand side of (\ref{identity}), now, over $\widetilde \lambda$. Using Euler's inversion formula for the gamma function, the previous equation can be restated by the following composition,
\begin{equation}
\label{identity2}
\mathcal Q^{\lambda_1}_N(J|M|\lambda) = \mathcal Q_3^{\lambda_1}(j_1,j_2,-1-j|m_1,m_2,-m|\lambda_2,\lambda') \mathcal Q_{N-1}^{\lambda'}(j,j_3,\dots,j_N|m,m_3,\dots,m_N|\widetilde \lambda).
\end{equation}

A suitable value of $\lambda_1$ that greatly simplifies the calculations when using these expressions is $\lambda_1=j_1\pm m_1$, since for these values all correlation functions appearing in the right hand side on (\ref{bbfinal2b}) have a lowest or a highest-weight state in the first insertion, respectively. When choosing so we shall omit any explicit reference to $\lambda_1$ and we will use $M^{\pm}=(m_2,\dots,m_N)$ instead of $M$ in the correlators, namely, we shall write
\begin{equation}
\label{bbfinal2b345}
\Phi_N(J|M|Z) = \mathcal Q_N(J|M^{\pm}|\lambda) \Phi_N(J|M^{\pm}\pm \lambda|Z).
\end{equation}
In general, we shall use the superscript ``$\pm$'' in order to indicate that the quantum numbers are consistent with a lowest or highest-weight state, respectively, in the first insertion point.

Let us apply (\ref{bbfinal2b345}) for the computation of the four-point function by replacing it in (\ref{eqfinal}) with an extremal-weight state in the first point. We obtain
\begin{equation}
\label{4ptupto}
\Phi_4(J|M|Z) =  \left| \Xi(\Delta|Z) \right|^{-2} \int_{\mathcal C^+} dj \, \mathcal Q_4(J|M^{\pm}|\lambda) \mathcal D_j(J|M^{\pm}\pm \lambda) {\mathcal G}_j (J|M^{\pm}\pm \lambda|z).
\end{equation}
If this expression actually gives the correct factorized form for the four-point function, the dependence of the integrand for the leading term as $z \rightarrow 0$ must go as $|z|^{2\left(\Delta_j-\Delta_{j_1}-\Delta_{j_2}\right)}$, so that $\mathcal Q_4(J|M^{\pm}|\lambda)$ should act trivially if contracted with this factor, and the remaining coefficient should equal $\mathcal D_j(J|M)$. While the first assertion straightforwardly follows from (\ref{confblock}), the second one is not so evident. In order to prove it we need to compute $\mathcal Q_4(J|M^{\pm}|\lambda) \mathcal D_j(J|M^{\pm}\pm \lambda)$ by replacing (\ref{cjm}) and (\ref{bbfinal2bbis}) in it. If we set $m \rightarrow m'=m \mp \lambda_1 \pm \lambda_2$ in (\ref{cjm}) and, instead of (\ref{bbfinal2bbis}), we use (\ref{identity2}) with $N=4$ while writing $\mathcal Q_4(J|M^{\pm}|\lambda)$, it is easy to realize that the operator $\mathcal Q_{3}^{\lambda'}(j,j_3,j_4|m,m_3,m_4|\lambda_3,\lambda_4)$ appearing in this decomposition acts non-trivially only on the last three-point function in (\ref{cjm}), namely, $\Phi_3(j,j_3,j_4|m',m_3\pm\lambda_3,m_4\pm\lambda_4|0,1,\infty)$, giving, by virtue of (\ref{bbfinal2b}), $\Phi_{34}(j|m)$. Since this function does not depend on $\lambda_2$, we have
\begin{eqnarray}
\label{identity3}
&& \mathcal Q_4(J|M^{\pm}|\lambda) \mathcal D_j(J|M^{\pm}\pm \lambda) = \nonumber \\
&&  \left(\mathcal Q_3(j_1,j_2,-1-j|m_2,-m|\lambda_2,\lambda') \Phi_{3}(j_1,j_2,-1-j|m_2\pm\lambda_2,-m'|0,1,\infty) \right) \Phi_{34}(j|m),
\end{eqnarray}
where we have used that $\Phi_{12}(j|m) \Phi_2(j|-m)^{-1}=\Phi_{12}(-1-j|m)$ for any value of $j$ and $m$, an identity proved in \cite{Teschner:1999ug}. The parenthesis in (\ref{identity3}) reduces to $\Phi_{3}(j_1,j_2,-1-j|m_1,m_2,-m|0,1,\infty)$, and, therefore, 
\begin{equation}
\label{ident}
\mathcal Q_4(J|M^{\pm}|\lambda) \mathcal D_j(J|M^{\pm}\pm \lambda)=\mathcal D_j(J|M),
\end{equation}
as we wanted to prove.

Concerning the conformal blocks, we have
\begin{equation}
\label{confblo22}
{\mathcal G}_j (J|M|z) = G_j(J|M)^{-1} \mathcal Q_4(J|M^{\pm}|\lambda) G_j(J|M^{\pm}\pm \lambda) {\mathcal G}_j (J|M^{\pm}\pm \lambda|z).
\end{equation}
Indeed, from (\ref{cjm}), (\ref{confblock}) and (\ref{ident}), it is likely that $\mathcal Q_4(J|M^{\pm}|\lambda)$ intertwines between the generating operators ${\mathcal K}_j (J|M|z)$ and ${\mathcal K}_j (J|M^{\pm}\pm \lambda|z)$, namely,
\begin{equation}
\label{confblock22bbb}
{\mathcal K}_j (J|M|z) \mathcal Q_4(J|M^{\pm}|\lambda) = \mathcal Q_4(J|M^{\pm}|\lambda) {\mathcal K}_j (J|M^{\pm}\pm \lambda|z).
\end{equation}

Interestingly enough, since $\mathcal Q_4(J|M^{\pm}|\lambda)$ does not depend on $z$, the descendant contributions to the conformal blocks of a given level can be related through (\ref{confblock22bbb}) to those contributions with a lowest or a highest-weight state coming strictly from the descendants of the very same level, {\em i.e.}, identity (\ref{confblock22bbb}) is realized order by order when formally expanding the generating operators in powers of $z$.

\section{The spectral flowed four-point function}
\label{sec4}

The standard method for computing correlation functions violating the spectral flow number conservation was exhaustively discussed in \cite{Maldacena:2001km,Fateev}, and it was used in these references and in \cite{Iguri:2007af} in order to compute three-point functions with a single unit of spectral flow. It requires the introduction of al least one spectral flow operator for each unit of spectral flow involved, followed by an ad-hoc procedure for removing the dependence of the resulting expression on the unphysical points in which these operators were originally inserted. Although this method is clear from a strictly theoretical point of view, dealing with such a large number of insertions makes it useless when trying to compute $N$-point correlators with $N$ greater than $3$.

In this section we shall consider the $AdS_3$-WZNW model four-point function violating the spectral flow number conservation in a single unit, namely, we will study $\widehat{\Phi}_4(J|M|Z)$. We will bypass the problem of having extra insertions by exploiting the series identification, following arguments similar to those used in \cite{Giribet:2006eb} in order to validate some expressions previously found in \cite{Maldacena:2001km} for the spectral flowed three-point function with one extremal-weight state.

As we have already pointed out, the series identification is an isomorphism between a spectral flow image of a representation induced by a lowest-weight module with spin $j$ and the image of a representation induced by a highest-weight module with a spin $j'=-k/2-j$ with one additional unit of spectral flow. The series identification generally maps flowed primary fields to flowed descendants with the unique exception of the spectral flowed extremal-weight states, that are mapped the one to the other. If the normalization of the spectral flowed states coincides with the unflowed ones, namely, if the two-point function involving two flowed operators equals the propagator (\ref{2ptm}) of the associated unflowed state, we have
\begin{equation}
\label{seriesid1}
\Phi_{j}^{\omega-1}(-j|z) = \frac{\pi^2}{B(j)} \Phi_{j'}^{\omega}(j'|z).
\end{equation}
The consistency of this equality can be easly checked within the context of two and three-point correlators. It follows form (\ref{seriesid1}) that
\begin{equation}
\label{seriesid3}
\Phi_{j}^{\omega+1}(j|z) = \frac{B(j')}{\pi^2} \Phi_{j'}^{\omega}(-j'|z),
\end{equation}
thus, if we set $\omega=\mp 1$, we can write
\begin{equation}
\widehat{\Phi}_{j}(\mp j |z) \propto \Phi_{j'}(\pm j'|z),
\end{equation}
the proportionality constant being those in (\ref{seriesid1}) or (\ref{seriesid3}), respectively.

Under this identification we have
\begin{equation}
\label{identif}
\widehat{\Phi}_4(J|M^{\pm}|Z) \propto \Phi_4(J'|M^{\mp}|Z),
\end{equation}
and, therefore,
\begin{equation}
\label{4ptwprev}
\widehat{\Phi}_4(J|M^{\pm}|Z) \propto  \left| \Xi(\Delta'|Z) \right|^{-2} \int_{\mathcal C^+} dj \, {\mathcal D}_j(J'|M^{\mp}) {\mathcal G}_j (J'|M^{\mp}|Z),
\end{equation}
where, as before, the prime was introduced in order to indicate that the spin of the first insertion is $j'_1=-k/2-j_1$. Recall that the formula (\ref{4ptwprev}) is valid whenever
\begin{eqnarray}
\label{teschnerdomw1}
\left|\mbox{Re}\left(j_{21}^{\pm}\right)-1+k/2\right|<\frac{1}{2}, \quad \quad & j_{21}^+=1+j_1+j_2, \quad & j_{21}^-=j_1-j_2, \\
\label{teschnerdomw2}
\left|\mbox{Re}\left(j_{43}^{\pm}\right)\right|<\frac{1}{2}, \quad \quad & j_{43}^+=1+j_3+j_4, \quad & j_{43}^-=j_3-j_4.
\end{eqnarray}

We have $\Delta_{j'}=\widehat{\Delta}_{j,\mp j}$, so that $\Delta'=\widehat{\Delta}^{\pm}$ and $\Xi(\Delta'|Z)=\Xi(\widehat{\Delta}^{\pm}|Z)$. On the other hand, since $\widehat{\Phi}_3(J|M^{\pm}|Z) \propto \Phi_3(J'|M^{\mp}|Z)$ with the same proportionality constant as in (\ref{identif}), we have ${\mathcal D}_j(J'|M^{\mp})=\widehat {\mathcal D}_j(J|M^{\pm})$ where we are defining
\begin{equation}
\widehat {\mathcal D}_j(J|M) = \widehat \Phi_{12}(j|-m) \Phi_2(j|m)^{-1} \Phi_{34}(j|m).
\end{equation}
with $\widehat \Phi_{12}(j|m) = \widehat \Phi_3(j_1,j_2,j|m_1,m_2,m|0,1,\infty)$.

It follows that we can recast (\ref{4ptwprev}) as
\begin{equation}
\label{4ptwext}
\widehat{\Phi}_4(J|M^{\pm}|Z) = \left| \Xi(\widehat{\Delta}^{\pm}|Z) \right|^{-2} \int_{\mathcal C^+} dj \, \widehat{\mathcal D}_j(J|M^{\pm}) \widehat{\mathcal G}_j (J|M^{\pm}|z)
\end{equation}
where the conformal blocks, as before, take the form
\begin{eqnarray}
\label{confblockwnew}
\widehat{\mathcal G}_j (J|M^{\pm}|z) & = & |z|^{2\left(\Delta_j-{\Delta}_{j'_1}-\Delta_{j_2}\right)} {\mathcal D}_j(J'|M^{\mp})^{-1}{\mathcal K}_j(J'|M^{\mp}|z){\mathcal D}_j(J'|M^{\mp}) \nonumber \\
& = & \left|z^{\Delta_j-\widehat{\Delta}_{j_1,\mp j_1}-\Delta_{j_2}}\right|^2 \widehat{\mathcal D}_j(J|M^{\pm})^{-1}\widehat{\mathcal K}_j (J|M^{\pm}|z)\widehat{\mathcal D}_j(J|M^{\pm}),
\end{eqnarray}
and we have introduced
\begin{equation}
\label{posta}
\widehat{\mathcal K}_j (J|M^{\pm}|z) = {\mathcal K}_j(J'|M^{\mp}|z).
\end{equation}

We have thus found an integral domain (\ref{teschnerdomw1})-(\ref{teschnerdomw2}) in which a factorized form for the four-point function involving one singly spectral flowed extremal-weight state holds, the corresponding conformal blocks being generated by the unflowed generating operator with a modified spin. Our next task is to relax the extremal-weight state condition in (\ref{4ptwext}) as we already did in the unflowed case.

As it can be guessed from (\ref{confblockwnew}), the method we have used in the previous section must be modified when nontrivial spectral flow charges are involved. Indeed, the overall dependence on $z$ in (\ref{confblockwnew}) should be adjusted when changing the value of $m_1$ and, therefore, the operator relaxing the extremal-weight condition on the conformal blocks should be expected to depend on the worldsheet variables, unlike in the unflowed case. In the rest of this section we shall compute this operator.

The key point for obtaining the explicit form of such operator relies in the fact that the insertion of additional lowest or highest-weight vertices in (\ref{campbell}), respectively, has no effect on (\ref{bbfinal2b}) since $e^{\mp\alpha J_0^{\pm}} \Phi_{j}(\pm j|\xi) e^{\pm\alpha J_0^{\pm}} = \Phi_{j}(\pm j|\xi)$. It follows that, for any non-negative integer value of $W$,
\begin{eqnarray}
\label{bbidontbis}
&& \left\langle \prod_{i=1}^N \Phi_{j_i}(m_i|z_i) \prod_{a=1}^{W} \Theta^{\mp}(\xi_a) \right\rangle = \nonumber \\
&& ~~~~~ ~~~~~ \mathcal Q_N^{\lambda_1}(J|M^{\pm}|\lambda) \left\langle \Phi_{j_1}(m_1\mp\lambda_1|z_1) \prod_{i=2}^N  \Phi_{j_i}(m_i\pm \lambda_i|z_i) \prod_{a=1}^{W} \Theta^{\mp}(\xi_a) \right\rangle.
\end{eqnarray}
where we have introduced $\Theta^{\mp}(\xi)=\Phi_{-k/2}(\mp k/2|\xi)$.

According to \cite{Fateev}, the coset factor of $\Theta^{\mp}(\xi_a)$ acts as a spectral flow operator carrying a negative or a positive unit of spectral flow, respectively, therefore, both correlators appearing in this last equation can be identified, up to an overall worldsheet dependence, with correlation functions involving states in nontrivial spectral flow sectors. Indeed, the correlator in the left hand side, if divided by
\begin{equation}
F_N(J|\Omega|M|Z)\left|\prod_{i=1}^{N} \prod_{a=1}^{W} (z_i-\xi_a)^{\pm m_i} \right|^2 \prod_{a<b}^{W} |\xi_a-\xi_b|^{-k},
\end{equation}
can be identified with $\Phi_N(J|\Omega|M|Z)$, with $\omega=\mp W$, while the correlator appearing in the right hand side of (\ref{bbidontbis}) can, accordingly, be identified with $\Phi_N(J|\Omega|M\pm\Lambda|Z)$ once it is divided by
\begin{equation}
F_N(J|\Omega|M\pm\Lambda|Z)  \left|\left(\prod_{a=1}^{\omega} (z_1-\xi_a)^{\pm m_1 - \lambda_1} \right)\left(\prod_{i=2}^{N} \prod_{a=1}^{\omega} (z_i-\xi_a)^{\pm m_i+ \lambda_i}\right) \right|^2 \prod_{a<b}^{\omega} |\xi_a-\xi_b|^{-k}.
\end{equation}
Therefore, we have
\begin{equation}
\Phi_N(J|\Omega|M|Z) = \mathcal Q^{\lambda_1}_N(J|\Omega|M|Z|\lambda) \Phi_N(J|\Omega|M\pm\Lambda|Z),
\end{equation}
where $\lambda$ is, again, a multi-index for which Einstein convention is assumed to hold and the operator $\mathcal Q^{\lambda_1}_N(J|\Omega|M|Z|\lambda)$ is given by the composition
\begin{eqnarray}
\label{equati}
&& \mathcal Q^{\lambda_1}_N(J|\Omega|M|Z|\lambda) = F_N(J|\Omega|M|Z)^{-1} \left|\prod_{a=1}^{\omega} (z_1-\xi_a)^{-\lambda_1} \right|^2 \times \nonumber \\
&& ~~~~~ ~~~~~ ~~~~~ ~~~~~ \mathcal Q^{\lambda_1}_N(J|M|\lambda) \left|\prod_{i=2}^{N} \prod_{a=1}^{\omega} (z_i-\xi_a)^{\lambda_i}\right|^2 F_N(J|\Omega|M\pm\Lambda|Z).
\end{eqnarray}
This expression can be simplified if we exploit the fact that the dependence on the unphysical insertion points must vanish at the end, as it was stated in \cite{Fateev}, by setting $\xi_a,\bar \xi_a\rightarrow\infty$. Since $\sum_{i=2}^N\lambda_i=\lambda_1$, this limiting process leads to
\begin{equation}
\label{equati2}
\mathcal Q^{\lambda_1}_N(J|\Omega|M|Z|\lambda) = F_N(J|\Omega|M|Z)^{-1} \mathcal Q^{\lambda_1}_N(J|M|\lambda) F_N(J|\Omega|M\pm\Lambda|Z).
\end{equation}
Notice, as we have advanced in the previous section, that this operator actually depends on the worldsheet coordinates through $F_N(J|\Omega|M|Z)$ and $F_N(J|\Omega|M\pm\Lambda|Z)$, factors that prevent us for obtaining a factorized form for $\mathcal Q^{\lambda_1}_N(J|\Omega|M|Z|\lambda)$, analogous to (\ref{identity2}).

In order to apply (\ref{equati2}) to (\ref{4ptwext}), let $\Omega=(\mp 1,0,\dots,0)$. In this case, $F_N(J|\Omega|M|Z)$ simplifies further to
\begin{equation}
\widehat F_N(J|M|Z)=\left|\prod_{i=2}^N (z_1-z_i)^{\mp m_i}\right|^{2},
\end{equation}
so that,
\begin{equation}
\widehat F_N(J|M|Z)^{-1} \widehat F_N(J|M\pm\Lambda|Z) = \left|\prod_{i=2}^N (z_1-z_i)^{-\lambda_i}\right|^{2}.
\end{equation}
Therefore, for $\lambda_1=j_1\pm m_1$, we have
\begin{equation}
\label{expect}
\widehat \Phi_N(J|M|Z) = \widehat{\mathcal Q}_N(J|M^{\pm}|Z|\lambda) \widehat \Phi_N(J|M^{\pm}\pm\lambda|Z),
\end{equation}
with $\widehat{\mathcal Q}_N(J|M^{\pm}|Z|\lambda)$ given by
\begin{equation}
\label{equati4}
\widehat{\mathcal Q}_N(J|M^{\pm}|Z|\lambda) = \left| \frac{(-1)^{\lambda_1}\lambda_1!\Gamma(-2j_1)}{\prod_{i=1}^N \Gamma(-j_i\pm m_i)} \prod_{i=2}^N \frac{\Gamma(-j_i\pm m_i+\lambda_i)}{\lambda_i!(z_1-z_i)^{\lambda_i}} \right|^2.
\end{equation}
One last simplification takes place if we invoke the conformal invariance of $\widehat \Phi_4(J|M|Z)$. After replacing (\ref{4ptwext}), we have that (\ref{expect}) reduces to
\begin{equation}
\label{expect2}
\widehat \Phi_4(J|M|Z) = \left| \Xi(\widehat{\Delta}|Z) \right|^{-2} \int_{\mathcal C^+} dj \, \widehat{\mathcal Q}_4(J|M^{\pm}|z|\lambda) \widehat {\mathcal D}_j(J|M^{\pm}\pm\lambda) \widehat{\mathcal G}_j (J|M^{\pm}\pm\lambda|z),
\end{equation}
where $\widehat{\mathcal Q}_4(J|M^{\pm}|z|\lambda)$ equals $\widehat{\mathcal Q}_4(J|M^{\pm}|Z|\lambda)$ with $z_1=0$, $z_2=z=z_{12}z_{34}/z_{13}z_{24}$, $z_3=1$ and $z_4\rightarrow\infty$. A crucial point in doing this limit is that it restricts the sum implicit in (\ref{expect}) to run on those $3$-tuples with a vanishing last entry, {\em i.e.}, with $\lambda_4=0$, canceling, thus, all the dependence of $\widehat{\mathcal Q}_4(J|M^{\pm}|z|\lambda)$ on the fourth insertion. Indeed, we have that the action of $\widehat{\mathcal Q}_4(J|M^{\pm}|z|\lambda)$ equals the one of $\widehat{\mathcal Q}_3(j_1,j_2,j_3|m_2,m_3|0,z,1|\lambda_2,\lambda_3)$, namely,
\begin{equation}
\label{equati5}
\widehat{\mathcal Q}_4(J|M^{\pm}|z|\lambda) = \left| \frac{(-1)^{\lambda_1}\lambda_1!\Gamma(-2j_1)}{\prod_{i=1}^3 \Gamma(-j_i\pm m_i)} z^{-\lambda_2} \prod_{i=2}^3 \frac{1}{\lambda_i!}\Gamma(-j_i\pm m_i+\lambda_i) \right|^2.
\end{equation}

As before, if (\ref{expect2}) gives the correct factorized form for the spectral flowed four-point function, we should be able to check that the dependence of the integrand for the leading term as $z \rightarrow 0$ is the expected one and that the remaining coefficient equals $\widehat {\mathcal D}_j(J|M)$. Unlike in the unflowed case, where the corresponding operator had no effect on the cross-ratio dependence of the conformal blocks, here it does, and not all the terms in (\ref{equati5}) contribute to the leading order behavior of (\ref{expect2}). In fact, only the term with $\lambda_2=\lambda_1$ must be taken into account. Recalling from (\ref{confblockwnew}) that $\widehat{\mathcal G}_j (J|M^{\pm}\pm\lambda|z) \sim |z|^{2\left(\Delta_j-\widehat{\Delta}_{j_1,\mp j_1}-\Delta_{j_2}\right)}$ as $z\rightarrow 0$ it follows that the leading term of the integrand in (\ref{expect2}) is
\begin{equation}
\label{expect33}
\left| z^{\Delta_j-\widehat{\Delta}_{j_1,\mp j_1}-\Delta_{j_2}-\lambda_1} \frac{(-1)^{\lambda_1}\Gamma(-2j_1)\Gamma(-j_2\pm m_2+\lambda_1)}{\Gamma(-j_1\pm m_1)\Gamma(-j_2\pm m_2)} \right|^2 \widehat {\mathcal D}_j(J|m_2\pm\lambda_1,m_3,m_4).
\end{equation}
Notice that the dependence on $z$ is the expected one, since
\begin{equation}
\widehat{\Delta}_{j_1,\mp j_1}+\lambda_1 =\Delta_{j_1} - j_1 - \frac{k}{4} + \lambda_1 =\Delta_{j_1} \pm m_1 - \frac{k}{4} = \widehat{\Delta}_{j_1m_1}.
\end{equation}
On the other hand, we have from (\ref{equati4}) that 
\begin{equation}
\left| \frac{(-1)^{\lambda_1}\Gamma(-2j_1)\Gamma(-j_2\pm m_2+\lambda_1)}{\Gamma(-j_1\pm m_1)\Gamma(-j_2\pm m_2)} \right|^2 = \widehat{\mathcal Q}_2(j_1,j_2|m_2|0,1|\lambda_1),
\end{equation}
which equals $\widehat{\mathcal Q}_3(j_1,j_2,j|m_2,-m|0,1,\infty|\lambda_1,0)$ for any value of $j$ and $m$, and therefore
\begin{equation}
\left| \frac{(-1)^{\lambda_1}\Gamma(-2j_1)\Gamma(-j_2\pm m_2+\lambda_1)}{\Gamma(-j_1\pm m_1)\Gamma(-j_2\pm m_2)} \right|^2 \widehat \Phi_3(j_1,j_2,j|m_2\pm\lambda_1,-m|0,1,\infty)=\widehat \Phi_{12}(j|-m),
\end{equation}
showing that the coefficient in (\ref{expect33}) is $\widehat {\mathcal D}_j(J|M)$, as we wanted to prove.

It follows from (\ref{expect2}) that, as the external spins lie in (\ref{teschnerdomw1})-(\ref{teschnerdomw2}), the flowed four-point function $\widehat \Phi_4(J|M|Z)$ can be factorized as
\begin{equation}
\label{pruel}
\widehat \Phi_4(J|M|Z) = \left| \Xi(\widehat{\Delta}|Z) \right|^{-2} \int_{\mathcal C^+} dj \,  \widehat {\mathcal D}_j(J|M) \widehat{\mathcal G}_j (J|M|z),
\end{equation}
where, by virtue of (\ref{confblockwnew}),
\begin{eqnarray}
&& \widehat{\mathcal G}_j (J|M|z) = \nonumber \\
&& ~~ \left|z^{\Delta_j-\widehat{\Delta}_{j_1,\mp j_1}-\Delta_{j_2}}\right|^2\widehat {\mathcal D}_j(J|M)^{-1} \widehat{\mathcal Q}_4(J|M^{\pm}|z|\lambda)
 \widehat{\mathcal K}_j (J|M^{\pm}\pm\lambda|z)\widehat{\mathcal D}_j(J|M^{\pm}\pm\lambda).
\end{eqnarray}

Assuming that there exists an operator $\widehat{\mathcal K}_j (J|M|z)$ generating the conformal blocks in analo\-gy with (\ref{confblock}), the dependence of $\widehat{\mathcal Q}_4(J|M^{\pm}|z|\lambda)$ on $z$ would prevent it from being strictly an intertwining operator between $\widehat{\mathcal K}_j (J|M|z)$ and $\widehat{\mathcal K}_j (J|M^{\pm}\pm\lambda|z)$ but a ``twisted'' one, since in that case we would have
\begin{equation}
\label{pruel222}
\widehat{\mathcal K}_j (J|M|z) \widehat{\mathcal Q}_4(J|M^{\pm}|z|\lambda) = \left|z^{-\lambda_1} \right|\widehat{\mathcal Q}_4(J|M^{\pm}|z|\lambda) \widehat{\mathcal K}_j (J|M^{\pm}\pm\lambda|z).
\end{equation}

\section{The spectral flowed four-point correlator in the space-time picture}
\label{sec5}

Up to this point, we have obtained a domain (\ref{teschnerdomw1})-(\ref{teschnerdomw2}) in which the factorized expression (\ref{pruel}) for the four-point function of the $AdS_3$-WZNW model with one unit of spectral flow holds. This formula was obtained in the $m$-basis by assuming that one of the fields corresponds to the spectral flow image of a state lying in the discrete part of the unflowed spectrum.

As we have already pointed out, the $m$-basis is the more convenient one for introducing spectral flow charges, however, the space-time representation is the best suited for using the results within the context of the AdS/CFT conjecture, as the $x$-basis vertex operators serve as ingredients for the string theory operators describing states created by sources in the boundary of the target space. For example, if $\Phi_j(x|z)$ is a field associated with an unflowed state and $\Theta(z)$ is a spinless worldsheet vertex corresponding to the internal CFT, such that the sum of their scaling dimensions equals 1, the operator
\begin{equation}
V_{j}(x)\sim \int_{\mathbb C} d^2z \Phi_j(x|z) \Theta(z),
\end{equation}
can be interpreted as describing a string state created by a pointlike source located at $x$ on the boundary of $AdS_3$ and, by means of the AdS/CFT correspondence, it can be identified with a CFT operator at the same point. Scattering amplitudes involving operators in the space-time picture, unlike amplitudes with states in the $m$-basis as the ones we have computed in the previous sections, acquire a similar interpretation when integrated over the string worldsheet as correlation functions on the dual two-dimensional CFT.

For unflowed primary states, the definition of the coordinate basis vertex operators comes from the Euclidean model through analytic continuation and the corresponding correlators follow in the very same fashion from those of the $H_3^+$-WZNW model. When dealing with spectral flowed primary fields, the situation is more complicated since they generally lie in representations with energy unbounded below. A solution for this issue was proposed in \cite{Maldacena:2001km}. An arbitrary lowest-energy state can be seen from a spectral flowed frame with $\omega > 0$ as the lowest-weight state of a certain discrete representation of the global algebra generated by the zero-modes with a spin $K$ being equal to $-m-k\omega/2$. Similarly, if the flow number $\omega$ is negative, the spectral flow automorphism maps the same state into the highest-weight state of a discrete representation with $K=m+k\omega/2$. Since the algebra generated by $J_0^a$ is identified with the space-time isometries of the background and the global $SL(2)$ symmetries of the CFT at the boundary, the vertex operators having flowed primaries and their global descendants as moments were accordingly proposed in \cite{Maldacena:2001km} as those relevant for physical applications.

Note that the eigenvalues of ${J}^3_0$ and its antiholomorphic counterpart do not necessarily agree and, therefore, it will also be the case for the global right and left-moving $SL(2)$ spins, namely, spectral flowed vertex in the $x$-basis are no longer expected to be spinless operators, their space-time planar spin being given by the difference between $K$ and $\bar K$. This number has to be an integer in order for the corresponding correlation functions to be single-valued. On the other hand, since the lowest-weight state and the highest-weight both contribute to the same operator, flowed vertex are not labelled by the spectral flow number but its absolute value.

We shall denote the flowed vertex operators in the $x$-basis by $\Phi^{j\omega}_{K}(x|z)$, where $\omega$ is now the (positive) amount of spectral flow and the superscript $j$ was introduced in order to remind the spin of the unflowed states this vertex is built from. The corresponding correlators will be
\begin{equation}
\Phi_N^J(K|\Omega|X|Z)=\left\langle \prod_{i=1}^N \Phi^{j_i\omega_i}_{K_i}(x_i|z_i) \right\rangle,
\end{equation}
with $J$, $X$, $\Omega$ and $Z$ as before and $K=(K_1,\dots,K_N)$. As in the previous sections, when $w=1$ we shall write $\widehat \Phi^{j}_{K}(x|z)$ and, accordingly, the $N$-point function with $\Omega=(1,0,\dots,0)$ will be written as $\widehat \Phi_N^J(K|X|Z)$. Recall that in this case $K$ differs from $J$ just in its first entry, {\em i.e.}, $K=(K_1,j_2,\dots,j_N)$.

The transformation between the space-time picture and the $m$-basis is carried out in analogy with (\ref{fourier}), namely,
\begin{equation}
\label{fourierJM}
 \Phi^{j\omega}_{K}(N|z) = \int_{\mathbb C} d^2x x^{K+N} \bar x^{\bar K+ \bar N} \Phi^{-1-j,\omega}_{-1-K}(x|z),
\end{equation}
where $N$ is the eigenvalue of $J^3_0$, and, by means of this map, we have
\begin{equation}
\label{identif33}
 \Phi^{j\omega}_{K}(\pm K|z) = \Phi^{\mp\omega}_{j}(\pm K \pm k\omega/2|z).
\end{equation}

Based on the knowledge of the dependence of the amplitudes on the boundary coordinates, it was realized in \cite{Maldacena:2001km} that this last identity could be enough for determining the constants that remain unfixed after the invariance of the correlators under the global $SL(2)$ symmetry is invoked, {\em i.e.}, for transforming from the $m$-basis back to the space-time representation. This idea was successfully implemented in order to obtain the propagator for a state in an arbitrary spectral flow sector and the three-point function with two unflowed operators and a single vertex with $\omega=1$. Explicitly, it was proven that
\begin{eqnarray}
\label{2ptdef}
  && \Phi_2^J(K|\Omega|X|Z) = \left|z_{12}^{2\Delta^{j_1\omega_1}_{K_1}}\right|^{-2} \left|x_{12}^{2K_1}\right|^{2} V_{\mbox{\footnotesize conf}} \frac{|1+2K_1|^2}{\pi^2} \delta_{\omega_1\omega_2} \left[\delta(1+j_1+j_2) + \right. \nonumber \\
	&& ~~~~~ ~~~~~ ~~~~~ ~~~~~ ~~~~~ \left. B(j_1) \Lambda^{j_1\omega_1}_{K_1} \delta(j_1-j_2) \right],
\end{eqnarray}
where $K_1=K_2$,
\begin{equation}
\label{lambda}
   \Lambda^{j\omega}_{K}=c_{j,-1-K+k\omega/2} = \frac{\pi}{\gamma(-2j)} \frac{ \gamma_{\omega}\left(-1-j-K\right)}{\gamma_{\omega}\left(j-K\right)},
\end{equation}
with
\begin{equation}
   \gamma_{\omega}(x)=\gamma\left(x+k\omega/2\right),
\end{equation}
and
\begin{equation}
\Delta^{j_1\omega_1}_{K_1}=\Delta_{j_1}-\omega_1 K_1 - \omega_1 + k\omega_1^2/4.
\end{equation}
Notice that
\begin{equation}
   \lim_{K\rightarrow j} \Lambda^{j,\omega=0}_{K}=\frac{\pi^2}{V_{\mbox{\footnotesize conf}}(1+2j)^2},
\end{equation}
therefore, the regular term of the two-point function for unflowed states is reached in the same limit, namely, by taking 
\begin{equation}
\label{limit}
   \Phi_{j_i}(x_i|z_i)=\lim_{K_i\rightarrow j_i} \Phi^{j_i,\omega_i=0}_{K_i}(x_i|z_i),
\end{equation}
for $i=1,2$. The contact term is not expected to be obtained from (\ref{2ptdef}) since the global spins for both insertions agree.

Concerning the three-point function, it is given by
\begin{equation}
\label{3ptx1}
\widehat \Phi_3^J(K|X|Z) = |C(\widehat{\Delta}|Z)|^{-2}|C(K|X)|^2 \widehat D(J|K)
\end{equation}
where $\widehat{\Delta}=\left(\widehat \Delta^{j_1}_{K_1},\Delta_{j_2},\Delta_{j_3}\right)$ with $\widehat \Delta^{j_1}_{K_1}= \Delta^{j_1,\omega_1=1}_{K_1}$ and
\begin{equation}
\label{3ptx122}
\widehat D(J|K)=\frac{1}{\pi^2}\frac{\widehat D(J)  \gamma(1-j_1+K_1-k/2) \gamma(2+2K_1)}{ \gamma(2+K_1+j_2+j_3)\gamma(1+K_{31})\gamma(1+K_{12})}.
\end{equation}

We shall generalize these expressions for the spectral flowed four-point function with one vertex in the $\omega=1$ sector, namely, $\widehat \Phi_4^J(K|X|Z)$, following similar lines. We shall not be able to obtain any integral expression for this correlator outside the domain (\ref{teschnerdomw1})-(\ref{teschnerdomw2}), since beyond this region we are not allowed to equate its $m$-basis transform with (\ref{pruel}) by means of (\ref{identif33}). We will, therefore, assume that (\ref{teschnerdomw1})-(\ref{teschnerdomw2}) hold.

By virtue of (\ref{pruel}), we expect an overall dependence of $\widehat \Phi_4^J(K|X|Z)$ on $Z$ of the form $\left| \Xi(\widehat{\Delta}|Z) \right|^{-2}$ with $\widehat{\Delta}=\left(\widehat \Delta^{j_1}_{K_1},\Delta_{j_2},\Delta_{j_3},\Delta_{j_4}\right)$, the remaining worldsheet dependence being strictly given through the cross-ratio $z$. Concerning the normalization of the conformal blocks, it is reasonable to presume that it equals (\ref{product3pt}) but with the first unflowed structure constant replaced by its spectral flowed analogon, namely,
\begin{equation}
\label{product3ptflow}
\widehat{\mathcal D}_j(J|K) = \widehat D_{12}(j) B(j)^{-1} D_{34}(j),
\end{equation}
where $\widehat D_{12}(j) = \widehat D(j_1,j_2,j|K_1,j_2,j)$. Finally, in the factorization limit, the conformal block should correspond to a function reducing $\widehat \Phi_4^J(K|X|Z)$ to (\ref{pruel}) after transforming to the $m$-basis and using (\ref{identif33}). A choice that turns out to be appropriate is given by 
\begin{equation}
\widehat{\mathcal G}_j^J (K|X|z)\sim|z|^{2\left(\Delta_j-\widehat \Delta^{j_1}_{K_1}-\Delta_{j_2}\right)} G_j (K|X),
\end{equation}
with $G_j (K|X)$ defined as in (\ref{futGK})-(\ref{hyper}). Indeed, from (\ref{3ptx122}) it can be proven that
\begin{equation}
\label{3ptx122bis}
\widehat D(-1-J|-1-K) = \frac{ \widehat D(-1-J)\widehat W(J|M)}{W(K|\pm K_1,m_2,m_3)},
\end{equation}
with $m_1=\pm K_1 \pm k/2$ (see \cite{Cagnacci:2013ufa}, Eqs.~(70)-(71) for more details), so that the $m$-basis transform of $G_j (K|X)$, given by (\ref{minces}), when evaluated at a lowest or a highest-weight state and combined with (\ref{product3ptflow}), gives $\widehat {\mathcal D}_j(J|M)$, in agreement with (\ref{pruel}).

Summarizing, we have
\begin{equation}
\label{eucl4ptfloww}
\widehat \Phi_4^J(K|X|Z) = \left| \Xi(\widehat{\Delta}|Z) \right|^{-2} \int_{\mathcal C^+} dj \, \widehat{\mathcal D}_j(J|K) \widehat{\mathcal G}_j^J (K|X|z),
\end{equation}
expression that would be valid as long as $J$ satisfies (\ref{teschnerdomw1})-(\ref{teschnerdomw2}). As for the unflowed four-point correlator, we can extend the integration contour to $\mathcal C$ and write the following holomorphically factorized expression generalizing (\ref{eucl4ptfactb}),
\begin{equation}
\label{eucl4ptfactbanot}
\widehat \Phi_4^J(K|X|Z)  = \left| \Xi(\widehat{\Delta}|Z) \right|^{-2} \left|\Xi(K|X) \right|^2 \int_{\mathcal C} dj \, \widehat{\mathcal D}_j(J|K) \left| \widehat{\mathcal F}_j^J (K|x|z)\right|^2,
\end{equation}
with $\widehat{\mathcal F}_j^J (K|x|z)$ admitting a formal power expansion of the form
\begin{equation}
\widehat{\mathcal F}_j^J (K|x|z) = z^{\Delta_j-\widehat \Delta^{j_1}_{K_1}-\Delta_{j_2}} x^{-j+K_1+j_2} \sum_{n=0}^{\infty} z^n \widehat F_j^{J(n)}(K|x),
\end{equation}
where $\widehat F_j^{J(0)}(K|x) = F(-j+K_1-j_2, -j-j_3+j_4;-2j|x)$.

Two consistency checks can be performed on (\ref{eucl4ptfloww}). The first one is by means of (\ref{limit}). As $\widehat D(J|K)$ equals $D(J)$ in this limit, (\ref{3ptx1}) reduces to (\ref{3ptbasex}), and, accordingly, (\ref{eucl4ptfloww}) reduces to (\ref{eucl4pt}), as expected. Note that the domain (\ref{teschnerdomw1})-(\ref{teschnerdomw2}) is shifted to (\ref{teschnerdomain1})-(\ref{teschnerdomain2}) when setting $\omega_1=0$.

The second check follows from setting $j_2=-k/2$. In this case, $\widehat D_{12}(j) \propto \delta(j-j'_1)$, the proportionality constant being only dependent on $k$, and, therefore, the integration in (\ref{eucl4ptfloww}) can be straightforwardly performed. The expression obtained is in accordance with the particular four-point function computed in \cite{Minces:2005nb} using completely different techniques.

\section{Concluding remarks}
\label{sec6}

In this paper, starting with the $m$-basis expression of the unflowed four-point function with at least one state in the discrete spectrum of the $AdS_3$-WZNW model derived by analytic continuation from its Euclidean analogon, it was proven that the conformal blocks can be written in terms of those involving the extremal-weight state corresponding to the same series. The proof was based on a method already used in \cite{Satoh:2001bi,Becker:1993at,Iguri:2007af} for relaxing the extremal-weight state condition when computing some related structure constants.

A similar procedure was implemented for studying four-point functions with one unit of spectral flow. A factorized integral expression was firstly obtained for a correlator involving the spectral flow image of an extremal-weight state by means of the series identification and a more general correlation function was then derived by using the techniques of \cite{Satoh:2001bi,Becker:1993at,Iguri:2007af} adapted for dealing with nontrivial spectral flow charges.

The formulas were all transformed to the space-time representation, which is the best suited picture for a string theoretical interpretation of the results. The transformation back to the $x$-basis was performed following the ideas of \cite{Maldacena:2001km}. We have been able to determine the overall dependence of the singly flowed four-point function both on the worldsheet and the space-time coordinates, the normalization of the conformal blocks and their leading order behavior as the worldsheet cross-ratio tends to zero. Higher order contributions remain to be determined.

The space-time picture integral expression we have found for the spectral flowed four-point function has been obtained by assuming that the external spins belong to a specific domain in $J$-space. In order to deal with correlation functions with short as well as long string states, this formula should be analytically continued as it was done for the unflowed four-point correlator in \cite{Maldacena:2001km}, namely, by taking into account the additional discrete contributions coming from the poles crossing the integration contour while continuously varying the values of the spins beyond (\ref{teschnerdomw1})-(\ref{teschnerdomw2}). We hope to be able to address this and other related issues in the near future.

\section*{Acknowledgments}

We are grateful to A.~Solotar and C.~N\'u\~nez for carefully reading the manuscript. We thank M.~Gra\~na and all the people of IPhT (CEA/Saclay) for their hospitality. This work has been partially supported by the projects UBACyT 20020100100669, PICT-2012-0513 and PIP 11220110100005CO.


\end{document}